# Decoupled Temporal Encoding for Generative Recommendation

Pengfei Jia†*
Rajax Network Technology
(Taobao Shangou of Alibaba)
Beijing, China
pengfei.jiapengfei@alibaba-inc.com

Jingjian Wang*
Rajax Network Technology
(Taobao Shangou of Alibaba)
Beijing, China
jingjianwang.wjj@alibaba-inc.com

Jingmao Li
Department of Biostatistics
Yale School of Public Health
New Haven, United States
mr.lijingmao@gmail.com

Ge Zhang†
Rajax Network Technology
(Taobao Shangou of Alibaba)
Beijing, China
luoge.zg@alibaba-inc.com

Feng Shi
Rajax Network Technology
(Taobao Shangou of Alibaba)
Beijing, China
sam.sf@alibaba-inc.com

## Abstract

Positional encoding is a fundamental component of Transformer-based generative recommendation models, where user histories are modeled as autoregressive item sequences. Most positional encoding methods are inherited from natural language processing and mainly represent discrete item order. However, recommendation sequences go beyond ordered lists, as timestamps and temporal effects also shape item relations. Our work is motivated by a real-world food delivery and instant retail recommendation system, where user behavior exhibits multi-level temporal regularities, including recency effects, meal-time peaks, weekday-weekend shifts, and promotion-driven traffic bursts. Existing methods partially address this issue through timestamp features, interval embeddings, decay functions, or attention biases, but they usually inject heterogeneous temporal signals through a unified representation or a single modeling pathway, making it difficult to distinguish broad temporal dynamics from local order cues. To address this limitation, we propose **Decoupled Temporal Encoding (DTE)**, a lightweight framework for generative recommendation. DTE separates temporal dynamics from order information through two complementary modules: a personalized macro-temporal module that injects compact temporal primitives into item embeddings, and a time-gated micro-sequential module that introduces relative-order bias only when interactions are temporally dense. DTE is also parameter-efficient and deployment-friendly, allowing easy integration into existing systems. Offline experiments on public and industrial datasets demonstrate DTE's superior performance. Online A/B tests further show improvements of +1.8% Click-through Rate (CTR) and +3.0% Revenue Per Mille (RPM) with only +0.3% serving latency overhead over the baseline positional encoding scheme. DTE has been deployed to the full Shangou ad recommendation traffic in the Taobao App since January 2026, highlighting the practical value of decoupling macro-temporal context from sequential order in real-world generative recommendation.

†Corresponding author, *Equal contributions

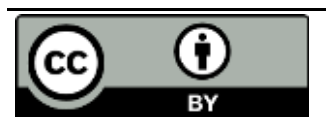

## 1 Introduction

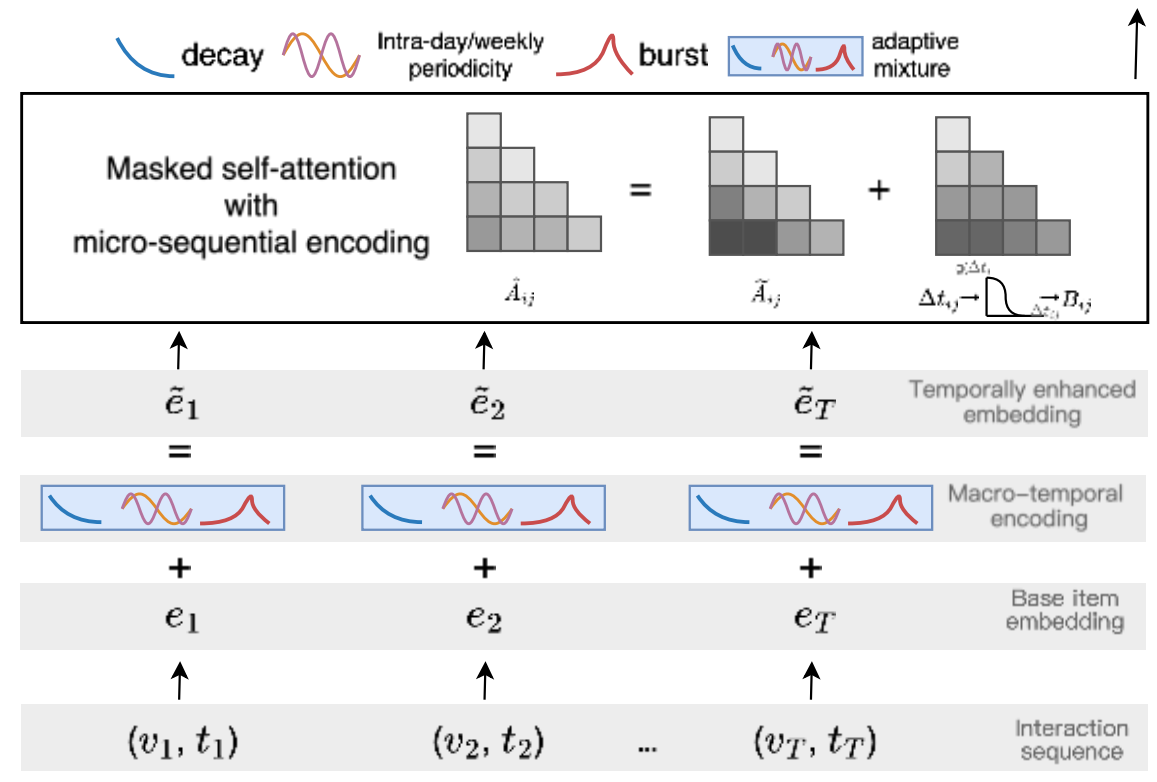

**Figure 1: Overview of DTE. The macro-temporal module models compact temporal context at the input level, while the micro-sequential module adds a time-gated order bias in self-attention based on pairwise time gaps.**

Generative recommendation has emerged as a powerful paradigm for modern recommender systems [5-10]. It represents user interaction histories as item sequences and reformulates next-item recommendation as an autoregressive sequence generation task. Compared with traditional candidate-scoring frameworks, this paradigm provides a natural way to model evolving user interests and has shown promising advantages in several real-world applications [5, 7, 10]. Similar to large language models (LLMs), Transformer architectures serve as a core building block of generative recommendation, where self-attention is used to capture complex dependencies among historical items. However, self-attention alone is permutation-invariant and does not explicitly encode sequence order, making positional encoding an indispensable component for Transformer-based generative recommendation [4]. Commonly used positional encoding schemes, such as sinusoidal embeddings [11] and Rotary Position Embedding (RoPE) [12], were originally developed for large language models (LLM) and are mainly designed to represent discrete token positions. In recommendation scenarios, however, user behavior sequences are not merely ordered lists: their timestamps contain rich continuous-time semantics that also shape item relations and user intent. These observations highlight the importance of temporal effects and motivate the development of time-aware positional encoding methods for generative recommendation.

Specifically, this work is motivated by, and initially designed for, the food delivery and instant retail recommendation system. In this practical scenario, user behavior exhibits multi-level temporal effects that are highly relevant to recommendation. First, recent interactions often carry stronger predictive value, reflecting a clear recency effect in user intent. Second, demand repeatedly concentrates around meal-related periods such as lunch and dinner, indicating strong intra-day periodicity. Third, user preferences and consumption needs may shift between weekdays and weekends, suggesting weekly temporal patterns: a user may frequently order prepared meals during weekday lunch hours, while showing weaker delivery demand or stronger grocery-oriented consumption on weekends. Finally, promotions, holidays, or operational events may induce short-term traffic surges, leading to bursty temporal deviations from regular behavior. These observations suggest that timestamps are not merely auxiliary context features, they encode multiple temporal effects that are essential for understanding evolving user intent. Meanwhile, we further highlight that such temporal effects complement, rather than replace, sequential order. Order information characterizes how user behaviors unfold. This distinction is especially important for temporally dense interactions, where relative order can still reveal meaningful item relationships. It is thus of importance here to explicitly incorporate both temporal and order effects when designing position encoding method.

To account for temporal effects, existing time-aware recommendation methods propose to inject time features, interval embeddings, or attention biases into sequential models [15–17]. While useful, such approaches often encode heterogeneous temporal effects in a unified representation, making it harder to distinguish broad temporal dynamics from local sequential order. This limitation is especially relevant in production recommendation, where temporal modeling must improve ranking quality under strict serving constraints. Large interval embedding tables can increase sparsity and maintenance cost, while heavier pairwise temporal parameterization can make deployment less convenient in existing ranking pipelines.

To address these challenges, we propose **Decoupled Temporal Encoding (DTE)**, a lightweight temporal- and order-aware position encoding framework for generative recommendation. DTE explicitly decouples "macro" temporal dynamics and "micro" sequential order effects at the architectural level by using two complementary modules. First, personalized macro-temporal module is injected at the input layer, compacting temporal primitives into item embeddings to model multi-level temporal effects. Second, a time-gated micro-sequential module introduces a relative-order bias at the attention level, allowing the model to leverage local order information when it is most informative. Through the gating mechanism, this order-aware bias is selectively activated for temporally dense interaction sequences, where local sequential order plays a particularly important role. By distinguishing broad temporal context from local sequential structure, DTE is well suited to food delivery, instant retail, and other real-world recommendation scenarios. Moreover, DTE is parameter-efficient, easy to integrate into existing Transformer-based ranking systems, and compatible with strict latency and memory constraints, thereby broadening its practical applicability in large-scale industrial recommender systems.

Offline experiments on both a public benchmark and a large-scale industrial dataset demonstrate the effectiveness of DTE. Importantly, we deploy DTE in the ranking stage of the Shangou advertisement recommendation system in the Taobao App, where it serves live production traffic under strict latency constraints. Online A/B tests show significant gains in key business metrics, including +1.8% CTR and +3.0% RPM, with only +0.3% average serving latency overhead. Given the promising advantages, DTE was rolled out to 100% of the full Shangou advertisement recommendation traffic in the Taobao App in January 2026.

Overall, our contributions are as follows:

- We identify a practical limitation of existing positional encoding methods in generative recommendation: rich temporal effects, such as recency and periodicity, are distinct from but complementary to local sequential order in modeling user intent, yet they are often conflated in a single representation
- We propose DTE, a lightweight temporal- and order-aware encoding framework that decouples "macro" temporal dynamics from "micro" sequential order via two complementary modules.
- We design DTE to be efficient and deployment-friendly. It requires only limited additional parameters, memory, and latency overhead, and enabling plug-and-play integration into existing Transformer-based generative recommendation systems.
- We validate DTE through comprehensive offline and online evaluations, including experiments on a public benchmark, a large-scale industrial dataset, online A/B testing, and full-traffic deployment in the Shangou advertisement recommendation system of the Taobao App.

**Code Availability**. The implementation of DTE is available at github.com/AlibabaResearch/DecoupledTemporalEncoding.

## 2 Related Work

### 2.1 Positional Encodings in LLM

Positional encoding is a fundamental component of Transformer-based LLMs. Existing methods can be broadly divided into absolute encodings, such as sinusoidal embeddings [11], and relative encodings, such as RoPE [12], T5-style relative bias [13], and Attention with Linear Biases (ALiBi) [14]. These methods are effective for representing discrete order and are widely used in sequence modeling. However, they are not explicitly designed to capture the continuous temporal patterns that appear in recommendation data.

### 2.2 Time-Aware Sequential Recommendation

A large body of work incorporates temporal information into sequential recommendation. Early methods such as TimeLSTM [15] introduced time-aware gating in recurrent models. Transformer-based methods such as BST [16] and TiSASRec [17] further injected time features or discretized time intervals into the model. More recently, bias-based approaches such as relative attention bias (RAB) [10] in HSTU have shown that attention scores can be modulated by lightweight context-dependent biases. These approaches improve temporal awareness, but they often represent multiple temporal effects in a unified space. As a result, it can be difficult to separate long-term temporal dynamics from sequential order. In addition, large interval embedding tables or more elaborate bias parameterizations can increase sparsity and make deployment less convenient in large-scale industrial systems.

### 2.3 Decomposed Sequence Modeling

Related work has also explored decomposition for long or heterogeneous sequences. Methods such as DSIN [2], SIM [3], TWIN [1], FIN [18] and DualGR [19] improve user modeling by decomposing session-level behavior or separating long-term and short-term interests. These approaches, however, typically operate at the level of sequence segmentation or interest aggregation. In contrast, DTE applies the decoupling principle directly to temporal encoding, separating continuous-time dynamics from sequential order in a lightweight plug-and-play manner.

## 3 Methodology

### 3.1 Problem Definition

Let $\mathcal{V}$ denote the item set. For a user $u$, we observe a chronological interaction sequence

$$S_u = \{(v_1, t_1), (v_2, t_2), \dots, (v_T, t_T)\}, \tag{1}$$

where $v_i \in \mathcal{V}$ is the interacted item and $t_i$ is its timestamp. Given the historical sequence, the goal is to predict the next item $v_{T+1}$.

Following the generative recommendation paradigm of HSTU [10], we use a decoder-only Transformer architecture to causally model the history sequence, and employ a lightweight task tower to match this representation with target items. Let $\boldsymbol{e}_i \in \mathbb{R}^d$ denote the base item embedding of $v_i$. Standard positional encoding injects order information into the model by modifying the input embeddings or by introducing relative bias into attention.

In a standard self-attention layer, the attention score between positions $i$ and $j$ is computed as

$$\boldsymbol{A}_{ij} = \frac{\boldsymbol{Q}_i \boldsymbol{K}_j^\top}{\sqrt{d}}, \tag{2}$$

where $\boldsymbol{Q}_i$ and $\boldsymbol{K}_j$ are the query and key vectors at position $i$ and $j$, respectively, obtained from the input embeddings through learned linear projections. A causal mask is applied so that each position can only attend to its previous positions. DTE builds on this backbone attention form by introducing temporal and order modifications at two complementary levels: macro-temporal information is injected into the input representations, while micro-sequential order is incorporated via a time-gated attention bias.

### 3.2 Design Principle

User behavior sequences often exhibit multi-level temporal effects. Recency and periodicity provide broad temporal context and are often user-dependent. For example, a user may repeatedly place orders around weekday lunch hours, or show stronger preferences for retail categories on weekends. These patterns suggest that timestamps contain important semantic information. However, temporal information alone is not sufficient to fully characterize a behavior sequence. When multiple interactions occur within a short time window, their timestamps or time gaps may provide limited distinction. In this case, their relative order can help avoid ambiguity. Therefore, our target is to design an effective positional encoding framework that jointly considers temporal effects and sequential order.

DTE proposes to use one mechanism for macro-temporal context and another for sequential order when timestamps become too dense to be informative. As illustrated in Figure 1, this leads to two complementary modules: (1) macro-temporal encoding that injects compact temporal context into input embeddings, and (2) micro-sequential encoding that adds a time-gated relative-order bias in attention. This decoupled design avoids forcing heterogeneous temporal effects into a single representation and also reflects practical deployment trade-offs: in production recommendation, temporal encoding must improve ranking quality without introducing noticeable latency or memory overhead, while remaining lightweight and easy to integrate into an existing Transformer ranking pipeline.

### 3.3 Macro-Temporal Encoding

The macro-temporal module is designed to capture common long-range temporal effects in a compact and interpretable way. Instead of using a large interval embedding table, we represent macro-temporal information with a small set of temporal primitives.

For each interaction $(v_i, t_i)$, we define the temporally enhanced embedding $\tilde{\boldsymbol{e}}_i$ as:

$$\tilde{\boldsymbol{e}}_i = \boldsymbol{e}_i + \boldsymbol{r}\, m(t_i), \tag{3}$$

where $m(t_i)$ is a scalar temporal signal, and $\boldsymbol{r} \in \mathbb{R}^d$ is a learnable projection vector that maps the scalar into the embedding space. In this way, the temporal signal remains compact while its effect is aligned with the item embedding space. We model $m(t_i)$ as an adaptive mixture of four temporal components:

$$m(t_i) = \beta_\Delta e^{-\lambda \Delta t_i} + \beta_h h(t_i) + \beta_w w(t_i) + \beta_p p(t_i), \quad (4)$$

where $\Delta t_i$ is the time gap between $t_i$ and the current prediction time. $\lambda \geq 0$ is a learnable decay coefficient. $\beta_\Delta$, $\beta_h$, $\beta_w$, and $\beta_p$ are mixture weights produced by a lightweight gating network:

$$[\beta_\Delta, \beta_h, \beta_w, \beta_p] = \text{softmax}(\text{MLP}(\boldsymbol{z}_u)), \quad (5)$$

where $\boldsymbol{z}_u$ denote the feature vectors for user. In practice, we compute $\boldsymbol{z}_u$ by average pooling over the historical item embeddings. $\text{MLP}(\cdot)$ denotes a lightweight multilayer perceptron that maps $\boldsymbol{z}_u$ to the logits. Eq. (5) allows the model to adapt the importance of temporal primitives to each user while keeping the temporal signal compact. For example, the model may place more weight on intra-day and weekly periodicity for users with repeated weekday lunch orders. In contrast, recency component may receive more weight for users with less regular but more recent interactions. This design allows DTE to model personalized temporal patterns without introducing heavy user-specific temporal parameters.

The four components are defined as:

- Recency decay:

$$e^{-\lambda \Delta t_i}. \quad (6)$$

- Intra-day periodicity:

$$h(t_i) = \sin\left(2\pi \frac{\text{hour}(t_i)}{24}\right). \quad (7)$$

- Weekly periodicity:

$$w(t_i) = \sin\left(2\pi \frac{\text{weekday}(t_i)}{7}\right). \quad (8)$$

- Burst-sensitive variation:

$$p(t_i) = \log\left(1 + \text{ReLU}\left(\frac{x_{t_i} - \bar{x}_{t_i}}{\bar{x}_{t_i} + \epsilon}\right)\right), \quad (9)$$

  where $\epsilon$ is a small constant for numerical stability, $x_{t_i}$ denotes a traffic-volume statistic of the same target entity in the corresponding time segment on the previous day, $\bar{x}_{t_i}$ denotes its 30-day historical average over corresponding time segments. In the industrial dataset, this entity is the target shop; in KuaiRand, it is the target video.

For these four components, recency decay in Eq. (6) follows a standard exponential decay formula. Intra-day and weekly periodicities are defined based on sinusoidal functions such that temporally adjacent points on the same cycle remain close, such as late night and early morning hours, or adjacent weekdays across a weekly boundary. The burst-sensitive variation in Eq. (9) captures short-term target-side traffic deviations from regular temporal patterns. It does not model intrinsic user burstiness directly. Instead, it provides target-side temporal context that complements user-level recency and periodicity. We therefore treat it as an optional, scenario-specific component whose benefit is larger in settings where such candidate-side traffic fluctuations are common and pronounced.

## 3.4 Micro-Sequential Encoding

Besides the macro-temporal encoding, DTE incorporates a micro-sequential encoding module to provide additional relative-order information. This module is particularly important when multiple interactions occur within a short time window, where timestamps or time gaps alone may provide limited distinction.

Specifically, DTE introduces a time-gated relative-order bias into self-attention as follows:

$$B_{ij} = -\alpha\, g(\Delta t_{ij})\, |i - j|, \quad (10)$$

where $\Delta t_{ij} = |t_i - t_j|$ is the absolute time gap between the two interactions. $\alpha \geq 0$ is a learnable coefficient. Function $g(\Delta t_{ij})$ is a soft gate that controls whether the order bias should be activated.

$$g(\Delta t_{ij}) = \sigma\left(\frac{\tau - \Delta t_{ij}}{\gamma}\right), \quad (11)$$

where $\sigma(\cdot)$ is the sigmoid function, $\tau$ is a learnable threshold defining the temporal range of dense interactions, and $\gamma$ is a fixed temperature parameter.

This design is intentionally conservative: sequential order is used as an auxiliary complementary information, rather than as a replacement, for timestamp information. It introduces relative-order information only when interactions are temporally dense, where timestamps alone may be insufficient to distinguish.

DTE then augments the backbone attention in Eq. (2) with the gated relative-order bias above. The attention score becomes

$$\widehat{\boldsymbol{A}}_{ij} = \widetilde{\boldsymbol{A}}_{ij} + \boldsymbol{B}_{ij}, \quad (12)$$

where $\widetilde{\boldsymbol{A}}_{ij}$ is the backbone attention computed from temporally enhanced inputs $\tilde{\boldsymbol{e}}_i$.

# 4 Experiments

## 4.1 Datasets and Experimental Setup

*4.1.1 Dataset.* We evaluate DTE on a large-scale industrial dataset and a public benchmark.

**Industrial**. We use a real-world dataset collected from Shangou advertisement recommendation in the Taobao App, which serves food delivery and instant retail. Each sample is an impression log with a binary click label, and the user history is constructed from interactions strictly preceding the impression timestamp within a one-year production window. All temporal features are computed causally from information available before the current sample time. For example, the time gap used in recency decay is the timestamp difference between the current prediction position and the preceding interaction, both taken from the logged history. Periodicity features are derived from event timestamps, and burst-sensitive signals from pre-aggregated past target-side traffic statistics. Figure 2 shows that the data exhibits strong temporal regularities. We set the maximum sequence length to 150 and truncate longer histories to the most recent interactions. We split the data chronologically by impression timestamp, using the earliest 85% of samples for training and the most recent 15% for evaluation.

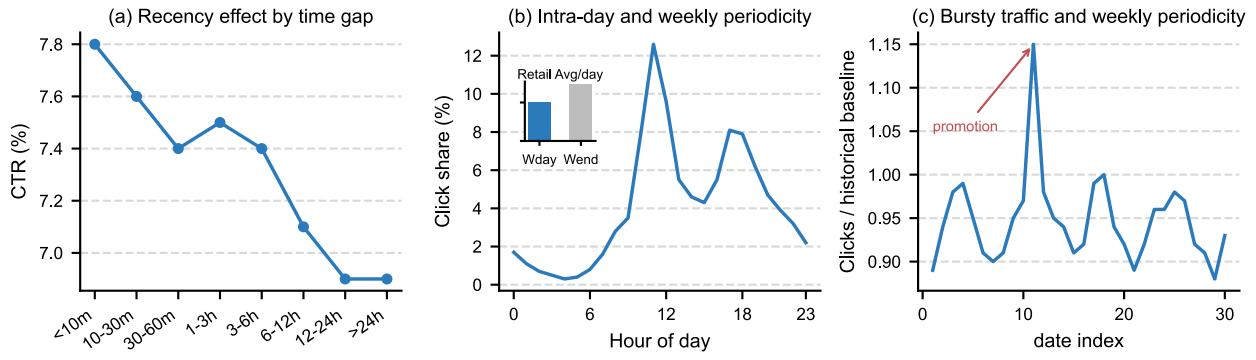

**Figure 2: Temporal characteristics of the industrial dataset. (a) CTR decays with interaction gap, indicating recency effects. (b) Clicks peak at meal times and vary between weekdays and weekends. (c) Shop traffic shows weekly patterns with bursts during promotions or events.**

**Table 1: Statistics of the datasets.**

| Dataset | Users | Item | Samples |
|---|---|---|---|
| Industrial | 400 M | 5 M | 2.5 B |
| KuaiRand 1K | 1000 | 4.4 M | 11.7 M |

**KuaiRand 1K** [20] is a public benchmark containing one month of user-video interaction logs. We formulate the task as impression-level click prediction with binary labels. We use the same history construction protocol and the same chronological split by impression timestamp as in the industrial dataset.

*4.1.2 Baselines.* We compare DTE with four groups of baselines: order-only positional encodings, including **Sinusoidal** [11], **RoPE** [12], and **ALiBi** [14]; time-aware sequential recommendation models, including **BST** [16], **TiSASRec** [17] and **RAB** [10]; strong baseline **BST + ALiBi**; and internal variants, including **w/o Macro**, **w/o Micro**, **w/o Recency**, **w/o Periodicity**, **w/o Burst**, **w/o Soft Gate**.

To ensure a fair comparison, all methods are evaluated under the same impression-level click prediction setting, with the same sequence construction protocol, chronological data split, training objective, and scoring framework. Whenever possible, the compared temporal encoding methods are implemented on top of the same decoder-only Transformer backbone used by DTE. For baselines originally introduced in different architectural or task settings, we retain their key temporal modeling components while adapting them to this shared production-oriented setup.

*4.1.3 Evaluation Metrics.* For offline evaluation, we use **GAUC** [21, 22] and **Logloss** as the main metrics on both datasets, **parameter**, **memory**, and **offline inference latency** overhead are also analyzed in Section 4.5. For online evaluation, we report **CTR**, **RPM**, and **latency**.

*4.1.4 Implementation Details.* Unless otherwise specified, all methods use the same decoder-only Transformer backbone (1 block, 8 attention heads, hidden size 64) for fair comparison and a lightweight MLP task tower (16-8-1) [10], optimized with Adam [23] (batch size 1024, learning rate 1e-4).

For DTE, the decay coefficient $\lambda$ and threshold $\tau$ are learned jointly with the model. The macro-temporal fusion weights are generated by an MLP (16-8-4) over an average-pooled user-history embedding, followed by softmax. In the micro-sequential module, pairwise time gaps are converted from milliseconds to seconds, and the gate temperature $\gamma$ is fixed to 10. Unless otherwise specified, hyperparameters are selected on the validation set.

For all learnable baselines, hyperparameters are selected on the validation set under the same tuning protocol, and the reported offline results correspond to the best validation checkpoint of each method. This setup is intended to isolate the effect of temporal encoding as much as possible, rather than differences in backbone architecture or task-specific scoring heads.

## 4.2 Offline Experiments

**Table 2: Offline SOTA comparison.**

| | | Industrial | | KuaiRand | |
|---|---|---|---|---|---|
| | | GAUC | Logloss | GAUC | Logloss |
| order-only | Sinusoidal | 0.6892 | 0.2628 | 0.8956 | 0.3016 |
| | RoPE | 0.6837 | 0.2634 | 0.9001 | 0.3008 |
| | ALiBi | 0.6861 | 0.2629 | 0.8994 | 0.3009 |
| time-aware | BST | 0.6956 | 0.2597 | 0.9096 | 0.2985 |
| | TiSASRec | 0.6970 | 0.2591 | 0.9094 | 0.2986 |
| | RAB | 0.6978 | 0.2593 | 0.9113 | 0.2981 |
| production baseline | BST + ALiBi | 0.6982 | 0.2589 | 0.9110 | 0.2979 |
| ours | DTE | 0.7098 | 0.2578 | 0.9243 | 0.2972 |

As shown in Table 2, DTE achieves the best performance on both datasets. Compared with order-only positional encodings, DTE consistently improves performance, suggesting that explicit temporal modeling is useful in recommendation sequences with rich timestamp information. Compared with time-aware baselines, it further improves performance, indicating that separating macro-temporal dynamics from sequential order can be more effective than combining all temporal effects in a single embedding. The consistent gains on both the industrial dataset and the public benchmark suggest that the benefit of decoupled temporal encoding is not limited to a single scenario.

## 4.3 Component Analysis

The ablation table 3 results show that both the macro and micro modules contribute to performance. Removing either module leads to a clear degradation, suggesting that long-range temporal context and sequential order are complementary. Removing the soft gate also hurts performance, which indicates that the sequential-order is more useful when activated selectively rather than uniformly. The burst term has a larger effect on the industrial dataset than on KuaiRand. This is consistent with the stronger entity-level traffic deviations in the industrial setting, where promotions more frequently create bursts relative to a shop's historical baseline. KuaiRand also provides analogous exposure-related statistics, but the corresponding deviations are weaker, so removing the burst term causes a smaller drop.

**Table 3: Ablation study.**

| | Industrial | | KuaiRand | |
|---|---|---|---|---|
| | GAUC | Logloss | GAUC | Logloss |
| Full DTE | 0.7098 | 0.2578 | 0.9243 | 0.2972 |
| w/o Macro | 0.6852 | 0.2631 | 0.8990 | 0.3008 |
| w/o Micro | 0.7006 | 0.2599 | 0.9101 | 0.2993 |
| w/o Recency | 0.6991 | 0.2601 | 0.9106 | 0.2986 |
| w/o Periodicity | 0.6975 | 0.2597 | 0.9088 | 0.2982 |
| w/o Burst | 0.7017 | 0.2605 | 0.9239 | 0.2975 |
| w/o Soft Gate | 0.6987 | 0.2588 | 0.9108 | 0.2980 |

## 4.4 Sensitivity and Robustness Analysis

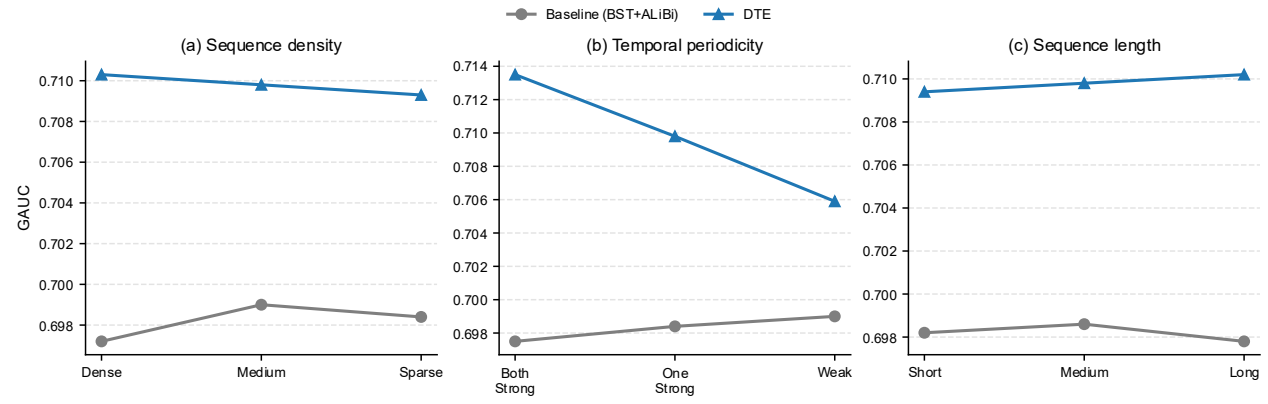

**Figure 3: DTE Sensitivity and robustness. (a) Larger gains on dense sequences where sequential ambiguity is severe. (b) Greater benefits for users with strong periodicity, validating the macro-temporal module. (c) Stable performance across varying history lengths.**

To further examine whether the gains of DTE align with its design motivation, we analyze model performance under different temporal conditions. Since DTE separates macro-temporal dynamics from sequential order, we study three representative factors: sequence density, temporal periodicity, and sequence length. All analyses in this subsection are reported using GAUC on the Industrial dataset.

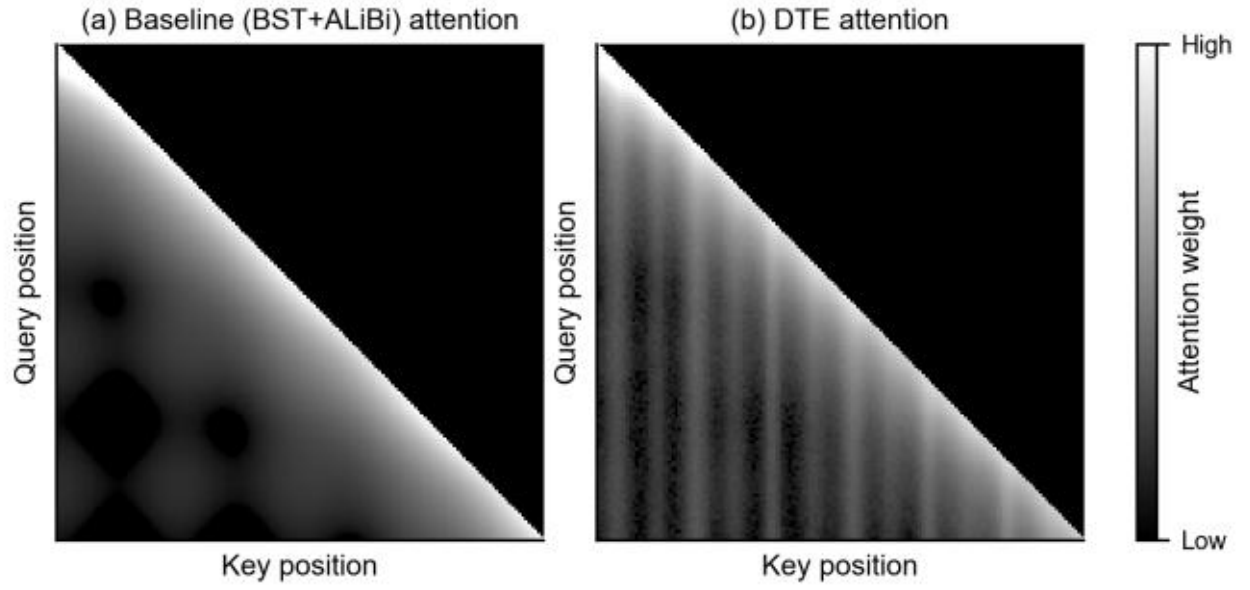

**Figure 4: Masked self-attention visualization (lighter = higher weight). (a) Baseline shows smooth recency decay concentrated near the diagonal. (b) DTE preserves causal recency while enhancing attention at periodic positions.**

*4.4.1 Sequence Density.* We first sort users by their average inter-event time gap and divide them into three equally sized groups: dense, medium, and sparse. This setting directly tests the micro-sequential module, which is designed to resolve sequential-order ambiguity when timestamps are too close to reliably determine event order.

As shown in Figure 3(a), DTE achieves the largest improvement on dense sequences, and the gain gradually decreases as the interaction gap becomes larger. This trend is consistent with our design: when timestamps are dense, the relative-order prior becomes more useful, whereas it becomes less important for sparse sequences where timestamps already provide sufficient temporal resolution.

*4.4.2 Temporal Periodicity.* We next measure each user's periodic behavior along two dimensions, intra-day and weekly. Based on these scores, users are divided into three groups: both strong, one strong, and weak. The both-strong group contains users whose histories exhibit clear daily and weekly regularities, while the one-strong group contains users with only one dominant periodic pattern. The weak group contains users whose behaviors are less regular over time. This analysis is intended to evaluate the macro-temporal module, which captures recurring behavioral rhythms via compact temporal primitives.

As shown in Figure 3(b), DTE yields larger gains for users with stronger periodicity, especially in the both-strong group. This indicates that the macro-temporal module is effective in capturing recurring time patterns, and that its benefit is more pronounced when user behavior follows regular temporal cycles. We attribute part of this gain to the user-adaptive weighting in Eq. (5), which allows the model to emphasize different temporal primitives for users with different routine strengths.

*4.4.3 Sequence Length.* Finally, we sort users by sequence length and divide them into three equally sized groups: short, medium, and long. This analysis is used to assess whether DTE remains stable when the amount of historical context changes.

Figure 3(c) shows that DTE consistently improves over the baseline across all history-length groups. The performance gap remains stable for both short and long histories, suggesting that DTE is not overly sensitive to the amount of available context and can generalize across sequences of different lengths.

*4.4.4 Attention Analysis.* To further examine whether the temporal components in DTE act in a compatible rather than conflicting manner, we visualize representative masked self-attention patterns in Figure 4. The baseline attention mainly exhibits smooth recency decay, with larger weights concentrated near the diagonal. After adding DTE, the attention map preserves the same causal recency structure but also shows repeated enhancement at periodic positions. This suggests that the macro-temporal primitives can reinforce periodic cues without disrupting the autoregressive ordering pattern. The resulting structure is consistent with the design of DTE: the macro module provides broad temporal context through the input embeddings, while the micro module only adds a lightweight sequential-order prior when interactions become temporally dense. Since burst-sensitive variation captures irregular short-term deviations, it is not expected to produce a regular geometric structure in the attention map.

## 4.5 Efficiency Analysis

We examine the efficiency of DTE by comparing its temporal modeling components against those of baseline methods, as all other architectural and optimization settings are shared. We therefore report three efficiency indicators: the number of additional trainable parameters, the corresponding FP16 parameter memory footprint, and offline inference latency under the same implementation setting.

Table 4 summarizes the overhead of each method. Fixed encodings such as Sinusoidal, RoPE, and ALiBi introduce no trainable parameters. In contrast, learned lookup-table methods incur substantially larger parameter and memory cost. With a maximum sequence length of 150 and embedding dimension of 64, BST-style embeddings require 9,600 additional parameters, or 18.75 KB in FP16. TiSASRec with 5,120 time-buckets requires 327,680 parameters, or 640.00 KB in FP16. RAB uses 32 relative-position buckets per attention head across 8 heads, resulting in 256 additional parameters, or 0.50 KB in FP16. By comparison, DTE introduces only 238 additional parameters, corresponding to 0.47 KB in FP16. Its offline inference latency is 3.02 ms, which remains close to other lightweight methods such as RAB (3.09 ms) and lower than BST and TiSASRec. Despite this compact footprint, DTE still models both macro-temporal context and micro-sequential order through a small gating MLP, a temporal projection vector, and a few scalar coefficients. These results confirm that DTE achieves a favorable balance between expressiveness and efficiency, consistent with the negligible online serving latency overhead observed in deployment.

## 4.6 Online A/B Test

**Table 4: Parameter, memory, and latency overhead.**

| | Params | FP16 Memory | Offline Inference Latency (ms) |
|---|---|---|---|
| DTE | 238 | 0.47 KB | 3.02 |
| BST | 9,600 | 18.75 KB | 3.18 |
| TiSASRec | 327,680 | 640.00 KB | 3.76 |
| RAB | 256 | 0.50 KB | 3.09 |

We deployed DTE in the ranking stage of the Shangou advertisement recommendation system in the Taobao App and evaluated it through a three-week online A/B test on live traffic. The experiment used user-level randomization with a 20% treatment bucket and a 20% control bucket under the same serving pipeline. Since the test lasted for three full weeks, it covered multiple complete weekly cycles and therefore reduced the risk that the observed gains were driven by short-term weekday–weekend fluctuations. Compared with the production baseline, DTE improved CTR by +1.8% and RPM by +3.0%, while increasing average serving latency by only +0.3% (15.4 ms to 15.5 ms). The improvements in CTR and RPM were statistically significant ($p < 0.01$, two-sided tests) and remained consistently positive via the test period. These results show that the temporal improvements brought by DTE translate into measurable online business gains with negligible serving overhead. Following the stable gains observed during the controlled experiment, DTE was subsequently rolled out to 100% of the main Shangou advertisement recommendation traffic in the Taobao App in January 2026. After the rollout, no systematic regression was observed in the monitored business or serving metrics, supporting its continued use in production.

## 5 Conclusions

DTE addresses a practical limitation of existing positional encoding methods in generative recommendation: heterogeneous temporal effects such as recency and periodicity are distinct from, yet complementary to, local sequential order, but are often conflated in a single representation. To resolve this, DTE decouples macro-temporal context from micro-level sequential order via two lightweight and deployment-friendly modules, requiring minimal additional parameters and latency overhead while enabling plug-and-play integration into existing Transformer-based ranking pipelines. This design achieves measurable gains in both offline ranking quality and online business metrics, and DTE was subsequently rolled out to 100% of Shangou advertisement recommendation traffic in January 2026. Its consistent improvements on the public benchmark further suggest that temporal decoupling is broadly applicable beyond this deployed setting.

Several lessons emerged from deployment. Recency and periodicity proved consistently useful across traffic conditions, while burst-sensitive variation is most effective when entity-level traffic exhibits strong deviations from historical baselines and should be treated as an optional scenario-specific component. DTE deliberately relies on compact temporal primitives and a soft-gated order bias rather than large interval embedding tables or heavier pairwise parameterization, trading some modeling flexibility for lower memory cost and easier integration into existing pipelines. One limitation is that DTE yields smaller gains for users with sparse histories or weak temporal periodicity, as temporal primitives provide limited signal in such cases. Future work could explore adaptive temporal encoding strategies that better handle low-activity users, or extend the decoupling principle to other sequential modeling scenarios.

## Acknowledgements

We sincerely thank the engineering team for their invaluable support in system implementation and online deployment.

## GenAI Usage Disclosure

The authors did not use any generative AI tools at any stage of this research, including code development, data analysis, or writing.